\documentclass[journal,onecolumn]{IEEEtran}

\usepackage{xcolor,soul,framed} 

\colorlet{shadecolor}{yellow}
\usepackage[pdftex]{graphicx}
\graphicspath{{../pdf/}{../jpeg/}}
\DeclareGraphicsExtensions{.pdf,.jpeg,.png}
\usepackage[cmex10]{amsmath}
\usepackage{mdwmath}
\usepackage{mdwtab}
\usepackage{eqparbox}
\newcolumntype{P}[1]{>{\centering\arraybackslash}p{#1}}
\usepackage{url}
\usepackage[caption=false]{subfig}
\usepackage{array, booktabs}
\usepackage{amsmath,amssymb,amsfonts}
\usepackage{mathtools}
\usepackage{multirow}
\usepackage{makecell}
\usepackage{algorithmic}
\usepackage[]{algorithm2e}
\usepackage{pifont}
\usepackage{verbatim}
\usepackage{subfig}
\usepackage{tabularx}
\usepackage{textcomp}
\usepackage{float}
\usepackage{cite}
\usepackage[normalem]{ulem} 
\usepackage{pifont}

\newcommand{\cmark}{\ding{51}} 
\newcommand{\xmark}{\ding{55}} 
\usepackage[T1]{fontenc}

\begin{document}

\title{Education 5.0: Requirements, Enabling Technologies, and Future Directions}


\author{
\IEEEauthorblockN{Shabir Ahmad\textsuperscript{1}\textsuperscript{*}, Sabina Umirzakova\textsuperscript{1}, Ghulam Mujtaba\textsuperscript{2}\textsuperscript, Muhammad Sadiq Amin\textsuperscript{3}, and Taegkeun Whangbo\textsuperscript{1}}
\thanks{S. Ahmed is with  Department of Computer Engineering, Gachon University, Gyonggi-do, Seongnam-si, Sujeong-gu, 113-120 Email: shabir@gachon.ac.kr}
\thanks{S. Umirzakova is with  Department of Computer Engineering, Gachon University, Gyonggi-do, Seongnam-si, Sujeong-gu, 113-120 Email: sabina@gachon.ac.kr}
\thanks{G. Mujtaba is with West Virginia University, Morgantown, USA 26505 Email: gmujtabakorai@gmail.com}
\thanks{M.S Amin is with Department of AI Software Engineering, Gachon University, Gyonggi-do, Seongnam-si, Sujeong-gu, 113-120 Email: engrsadiqamin@gmail.com}
\thanks{T. Whangbo is Department of Computer Engineering, Gachon University, Gyonggi-do, Seongnam-si, Sujeong-gu, 113-120 with Email: whangbotaegkeun@gmail.com}
}

\markboth{}
{Education 5.0: Requirements, Enabling Technologies, and Future Directions}



\IEEEtitleabstractindextext{%
\begin{abstract}
We are currently in a post-pandemic era in which life has shifted to a digital world. This has affected many aspects of life, including education and learning. Education 5.0 refers to the fifth industrial revolution in education by leveraging digital technologies to eliminate barriers to learning, enhance learning methods, and promote overall well-being. The concept of Education 5.0 represents a new paradigm in the field of education, one that is focused on creating a learner-centric environment that leverages the latest technologies and teaching methods. This paper explores the key requirements of Education 5.0 and the enabling technologies that make it possible, including artificial intelligence, blockchain, and virtual and augmented reality. We analyze the potential impact of these technologies on the future of education, including their ability to improve personalization, increase engagement, and provide greater access to education. Additionally, we examine the challenges and ethical considerations associated with Education 5.0 and propose strategies for addressing these issues. Finally, we offer insights into future directions for the development of Education 5.0, including the need for ongoing research, collaboration, and innovation in the field. Overall, this paper provides a comprehensive overview of Education 5.0, its requirements, enabling technologies, and future directions, and highlights the potential of this new paradigm to transform education and improve learning outcomes for students.
\end{abstract}

\begin{IEEEkeywords}
Education 5.0, personalized learning, adaptive learning, blended learning.
\end{IEEEkeywords}}

\maketitle

%
\IEEEdisplaynontitleabstractindextext

\section{Introduction}\label{sec:level1}
\IEEEPARstart{E}{}ducation is a basic human right, and delivering knowledge effectively has been a success metric for ages. Education systems have undergone extensive revamp over the past decade with the rise of the Internet of Things (IoT) and information communication technology (ICT). The integration of sensors and the processing of data through artificial intelligence (AI) technologies paved the way for the next-generation education systems \cite{al2020survey}. Education is linked to the industrial revolution, and the fourth industrial revolution has witnessed tremendous work towards the realization of Education 4.0 to enhance the learning experience through the use of ICT and IoT technology \cite{zhamanov2017iot,ur2016ict}. However, despite its many improvements over the conventional education systems, the demand for personalized tutoring and education and game-based learning envisioned a fifth revolution in Education.

Education 5.0 \cite{lantada2020engineering} is a futuristic term that aims to integrate advanced ICT technologies into the education system to enhance the learning experience and remove barriers to educating an individual. Thus, one of the fundamental goals of the Education 5.0 is to promote personalized learning, collaboration, and well-being through the use of digital tools such as AI, virtual reality (VR), and the IoT. Additionally, Education 5.0 focuses on developing 21st-century skills such as critical thinking, creativity, and problem-solving, rather than just rote learning and adds immersive experience in the classrooms using augmented reality and mixed reality applications \cite{mustafa2019immersive}. The ultimate goal of Education 5.0 is to create a more efficient, effective, and equitable education system that can adapt to the changing needs of society in the fifth industrial revolution.

\subsection{Market statistics and existing surveys}
\begin{figure}
\centering
\includegraphics[width=\textwidth, height=8cm]{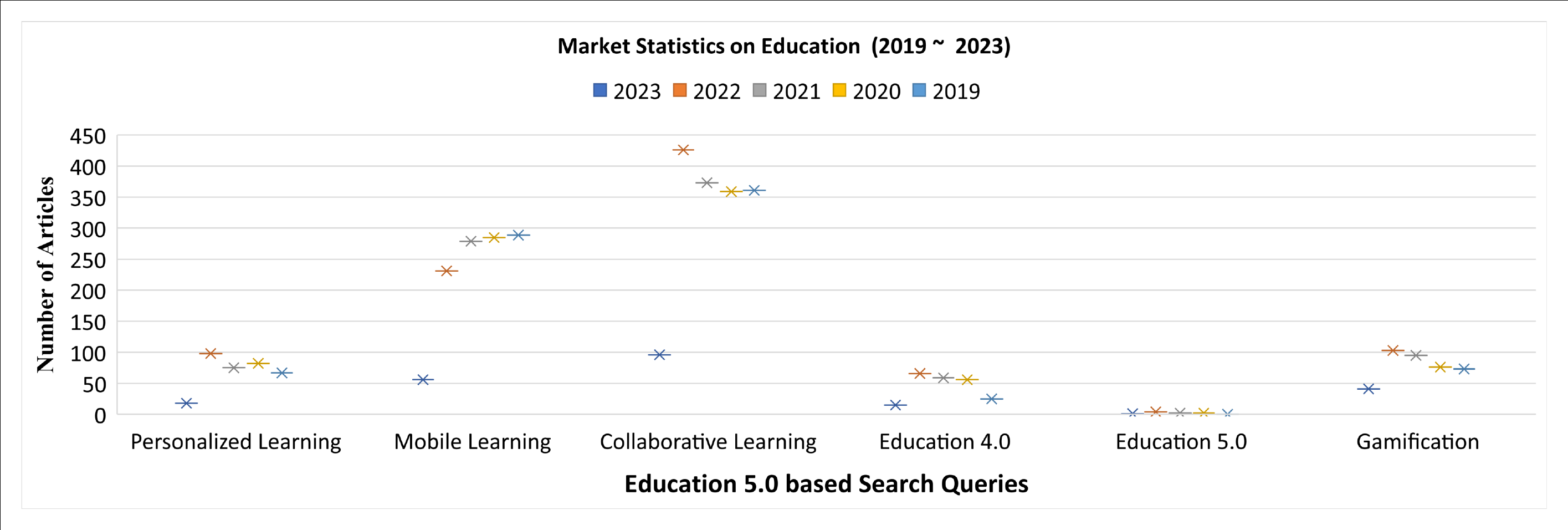}
\caption{\label{fig:evolution} Evolution of Education over the past years. The goal of the latest form of education has always been to use the state-of-the-art technology to aid in the learning process and improve students' engagement}
\end{figure}

\begin{figure}
\centering
\includegraphics[width=\textwidth]{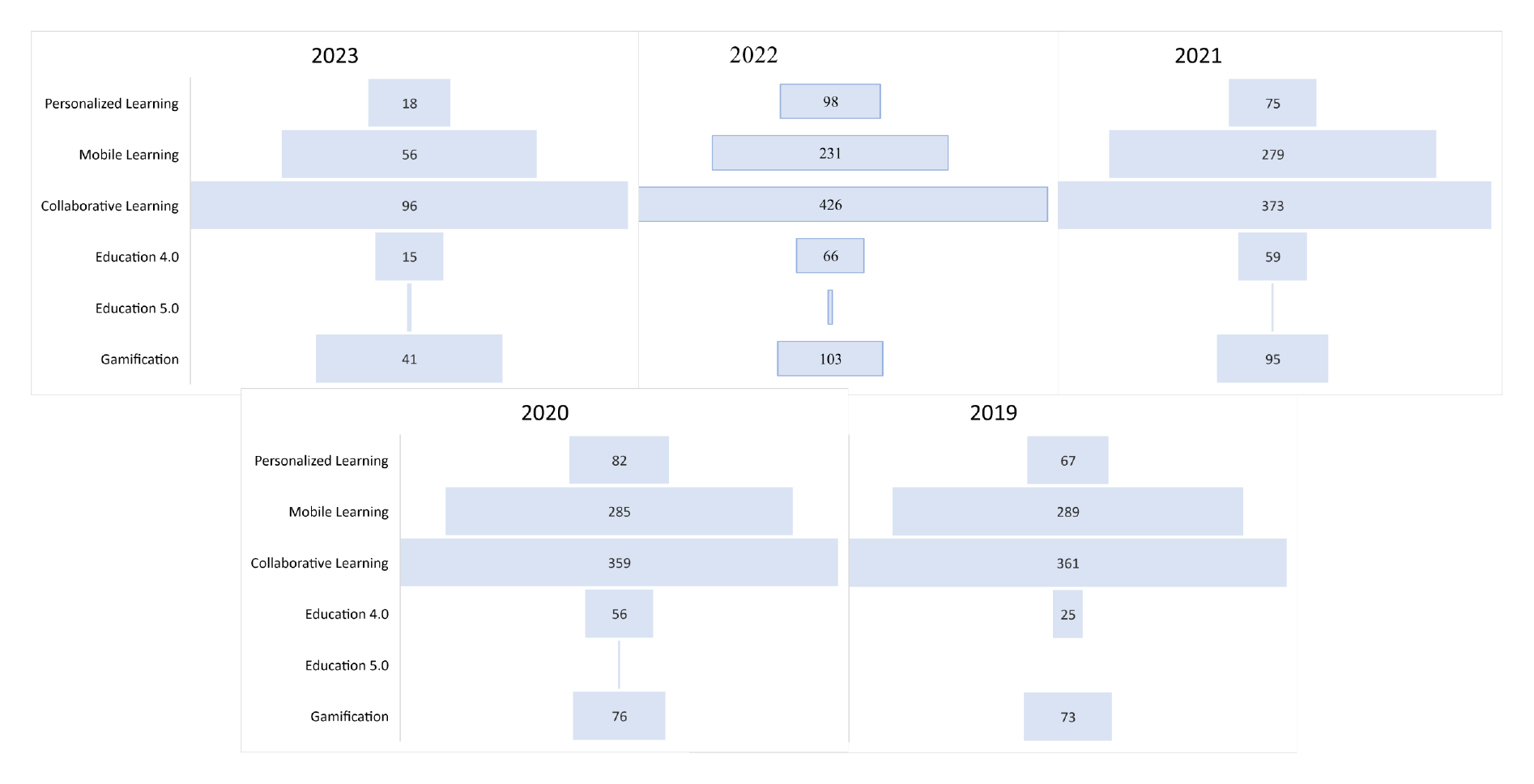}
\caption{\label{fig:evolution1} Evolution of Education over the past years. The goal of the latest form of education has always been to use state-of-the-art technology to aid in the learning process and improve students' engagement}
\end{figure}

According to a report by \cite{aston2022education}, the global education technology market is expected to grow at a compound annual growth rate of 17\% between 2021 and 2025. The report cites the growing demand for personalized learning and the increasing availability of digital content as driving factors behind this growth. Furthermore, a report by \cite{sharma2022artificial} predicts that the global e-learning market will reach \$374.3 billion by 2026, with a compound annual growth rate of 14.6\% from 2021 to 2026. This growth is attributed to the increased adoption of mobile and digital learning, the rising need for skill-based education, and the growing use of artificial intelligence in education. The concept of Education 5.0 is relatively new, so there is limited academic literature specifically focused on this approach to education. However, several articles discuss related topics, such as personalized learning, digital technology, and the use of data in education. For example, a study by Means, Toyama, Murphy, and Baki (2013) \cite{means2013} found that technology can support personalized learning by providing students with access to various resources and opportunities for feedback. Another study by  \cite{quadir2022analyzing} examined the use of data analytics in education and how it can be used to support personalized learning and improve student outcomes. A search on popular databases such as Google Scholar and Scopus using the keywords "Education 5.0" returns relatively only 9 records, indicating that this is a relatively new and emerging field. However, a search for related topics such as "personalized learning" and "digital education" returns a significantly large number of articles, as shown in Figure \ref{fig:evolution} and Figure \ref{fig:evolution1}. Therefore, though the idea of Education 5.0 centers around personalized education and collaborative learning has been around for the last decade, the notion of Education 5.0 as the next wave of Education is still a gap in the current state-of-the-art.

\begin{figure}
\centering
\includegraphics[width=\textwidth]{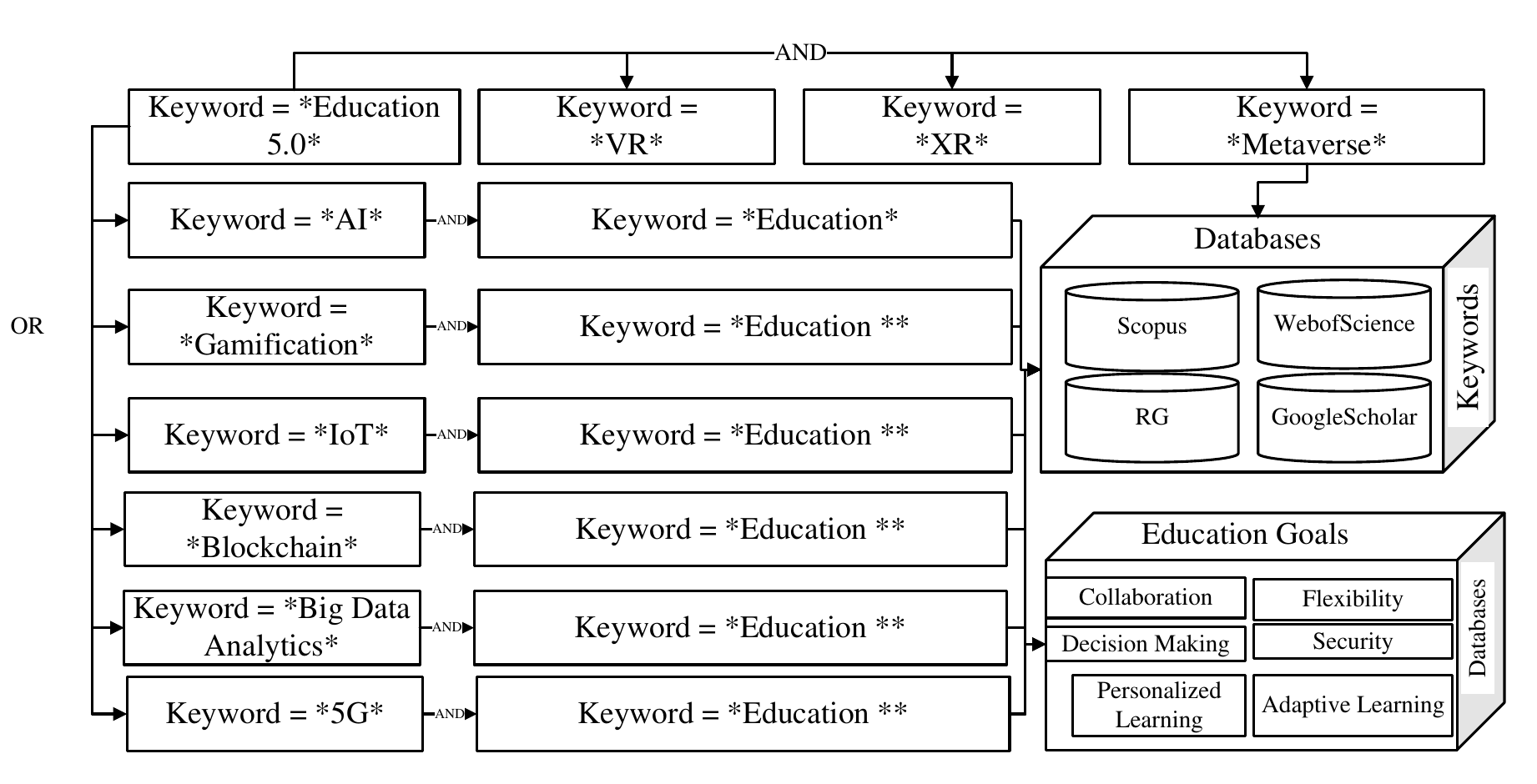}
\caption{\label{fig:data_evolution} Evolution of Education over the past years. The goal of the latest form of education has always been to use the state-of-the-art technology to aid in the learning process and improve students' engagement}
\end{figure}

There have been several surveys and tutorials published that address various aspects of education, including personalized learning, digital technology, and data analytics. Table \ref{table:attributes} highlights some of the studies which emphasize the need for one or some of the fundamental components in Education 5.0.

\begin{table}[t]
\centering
\caption{Education Attributes Addressed by Selected Surveys and Tutorials}
\label{table:attributes}
\begin{tabular}{|p{60pt}|p{125pt}|p{40pt}|p{35pt}|p{30pt}|p{45pt}|p{35pt}|p{45pt}|}
\hline
\textbf{Study} & \textbf{Contribution} & \textbf{Personalized Learning} & \textbf{Digital \newline Technology} & \textbf{Data  \newline Analytics} & \textbf{Collaborative \newline Learning} & \textbf{Mobile \newline Learning} & \textbf{Gamification} \\ \hline
Means et al. (2013) \cite{means2013} & Technology has the potential to support personalized learning & \xmark & \cmark & \xmark & \cmark & \xmark & \xmark \\ \hline
Quadir and Chen (2022) \cite{quadir2022analyzing} & Data analytics can be used to support personalized learning and improve student outcomes & \cmark & \cmark & \cmark & \cmark & \xmark & \xmark \\ \hline
Green et al. (2020) \cite{green2020fault} & Online learning management systems can be remodeled to use them more effectively & \xmark & \cmark & \xmark & \xmark & \xmark & \xmark \\ \hline
Nisha et al. (2022) \cite{raj2022systematic} & Recommender systems can be used to support personalized learning & \cmark & \cmark & \cmark & \xmark & \xmark & \xmark \\ \hline
Li et al. (2021) \cite{li2021design} & Mobile learning can support personalized learning in a blended learning environment & \cmark & \cmark & \xmark & \xmark & \cmark & \xmark \\ \hline
Major et al. (2021) \cite{major2021effectiveness} & Digital technology can improve access to education and support personalized learning & \cmark & \cmark & \xmark & \xmark & \xmark & \xmark \\ \hline
Greaney (2021) \cite{tapalova2022artificial} & Data analytics can be used to identify areas where personalized learning can be improved & \cmark & \xmark & \xmark & \cmark & \xmark & \xmark \\ \hline
Chen et al. (2020) \cite{chen2020csclrec} & Collaborative learning can support personalized learning & \cmark & \xmark & \xmark & \cmark & \xmark & \xmark \\ \hline
Ofosu et al. (2020) \cite{ofosu2020shift} & Shift towards gamification in education & \xmark & \xmark & \xmark & \cmark & \xmark & \cmark \\ \hline
Georgios et al. (2022) \cite{lampropoulos2022augmented} & Augmented reality and gamification to enable various learning techniques & \cmark & \xmark & \xmark & \cmark & \cmark & \cmark \\ \hline
\end{tabular}
\end{table}

\subsection{Motivation and Scope}

Education 5.0 covers a wide spectrum of attributes, such as using ICT and data analytics to make a more informed educational system. This survey proposes a two-dimensional framework for which the x-axis has different educational goals, such as personalized learning, collaborative learning and decision-making, while on the y-axis are the technologies needed to pursue a particular goal, such as Big data analysis, IoT, and VR, to name a few. Thus, the scope of this survey is to review the literature on the above two-dimensional pairs and highlight their impact on overall education. Since these attributes are widely dispersed and inaccessible, this survey will make a foundation platform to propose the roadmap for realizing Education 5.0 by combining all these attributes and technologies.

\subsection{Survey organization}
The rest of this survey is organized as follows; Section 2 provides Background and foundation and discusses the evolution of Education over the years. Section 3 highlights some of the requirements. Section 4 summarizes the enabling technologies. Section 5 portrays the current state-of-the-art and future roadmap, section 6 identifies some crucial challenges, and section 7 concludes the paper.

\section{Background and Foundation}
\begin{figure}
\centering
\includegraphics[width=\textwidth]{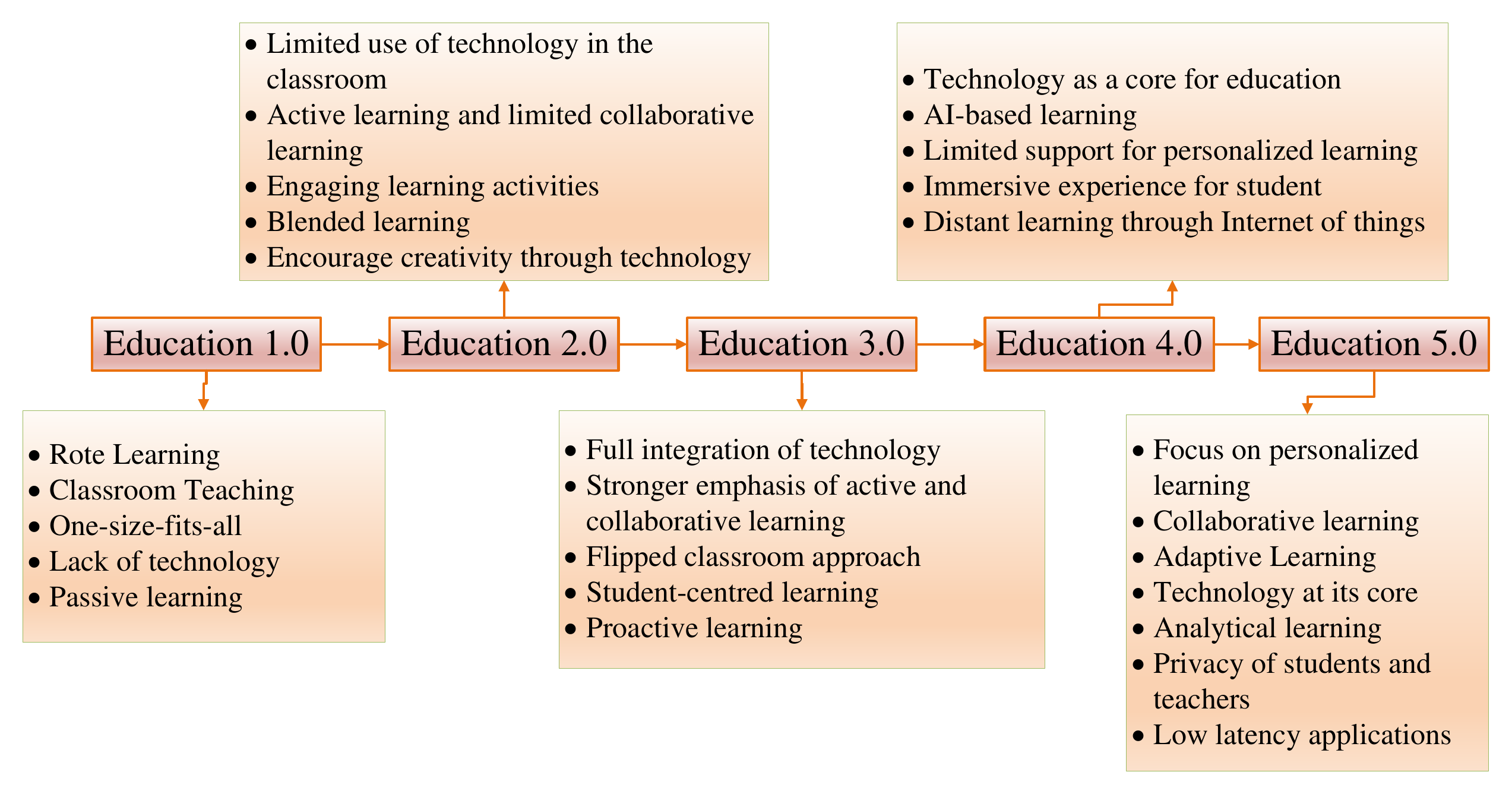}
\caption{\label{fig:schematic_evolution} Evolution of Education over the past years. The goal of the latest form of education has always been to use the state-of-the-art technology to aid in the learning process and improve students' engagement}
\end{figure}

In this section, we outline the background of the use of contemporary technologies to aid in elevating the quality of education is discussed to serve as a foundation for upcoming sections. The core foundation concepts are elucidated with the gradual evolution of Education from 1.0 to 5.0 in the subsequent sections. The evolution of Education 5.0 results from the advancements in technology and the changing needs of the workforce and society. It can be broken down into several stages, each building on the previous one and incorporating new technologies and approaches. The evolution of Education 5.0 was a long and lengthy process that incrementally added new approaches to support and assess the learning process and methodology. The schematic representation of this evolution is shown in figure \ref{fig:schematic_evolution}.

\subsection{Education 1.0}
Education 1.0, also known as traditional classroom teaching, is characterized by the teacher being the primary source of information and students being expected to memorize and repeat information. The primary focus of Education 1.0 was on rote learning, which is the memorization of information without understanding the underlying concepts \cite{Barkley2010}, \cite{Friedman2011}.

In Education 1.0, teaching methods were typically based on lecture-style instruction, and there was little use of technology in the classroom \cite{Warschauer2000}, \cite{Warschauer2002}. Students were expected to take notes, memorize information, and complete assignments and exams that were designed to test their ability to recall the information. The emphasis was on the teacher as the authority figure, and the students were expected to be passive learners \cite{Barkley2010}, \cite{Friedman2011}.

This type of education was often seen as a one-size-fits-all approach, and it did not take into account individual differences in learning styles and abilities \cite{Felder1988}, \cite{Felder2005}. The lack of technology also limited the resources and materials that were available to teachers and students \cite{Warschauer2000}, \cite{Warschauer2002}.

Several shortcomings in Education 1.0 led to the development of Education 2.0. Some of these include:

\textbf{Rote learning:} The primary focus of Education 1.0 was on rote learning, which is the memorization of information without understanding the underlying concepts \cite{Barkley2010}, \cite{Friedman2011}. This approach did not foster critical thinking or problem-solving skills, and it did not take into account individual differences in learning styles and abilities \cite{Felder1988}, \cite{Felder2005}.

\textbf{Lack of technology:} Education 1.0 had little use of technology in the classroom \cite{Warschauer2000}, \cite{Warschauer2002}, which limited the resources and materials that were available to teachers and students. This made it difficult for teachers to create interactive and engaging learning experiences, and it limited the opportunities for students to learn through new and innovative methods \cite{Warschauer2000}, \cite{Warschauer2002}.

\textbf{One-size-fits-all approach:} The traditional classroom teaching in Education 1.0 was a one-size-fits-all approach \cite{Felder1988}, \cite{Felder2005}, which did not take into account the individual needs of students. This made it difficult for teachers to provide personalized instruction and support to students with different learning styles and abilities \cite{Felder1988}, \cite{Felder2005}.

\textbf{Passive learning:} The emphasis in Education 1.0 was on the teacher as the authority figure, and the students were expected to be passive learners \cite{Barkley2010}, \cite{Friedman2011}. This approach did not encourage students to take an active role in their own learning, and it did not foster creativity or independent thinking \cite{Barkley2010}, \cite{Friedman2011}.

These shortcomings in Education 1.0 led to the development of Education 2.0, which introduced technology in the classroom and shifted the focus from rote learning to more active and collaborative learning\cite{garrison2009technology}. Education 2.0 also introduced the concept of using technology to enhance the learning experience and make education more accessible to all \cite{Warschauer2000}.

\subsection{Education 2.0}
Education 2.0 represents an evolution in the field of education, building upon the shortcomings of Education 1.0. Education 2.0 introduced technology in the classroom, and it shifted the focus from rote learning to more active and collaborative learning. One of the key features of Education 2.0 is the use of technology to create more interactive and engaging learning experiences. For example, the use of computers and the internet allowed for the use of digital resources and materials, such as videos, animations, and interactive simulations, which can be used to supplement traditional teaching methods and make learning more engaging and interactive \cite{dovcekal2015impact,ericsti2012teachers}.

Additionally, Education 2.0 introduced the concept of using technology to make education more accessible to all. With the internet and online resources, students can access a wealth of information and educational materials, regardless of their location or socio-economic background \cite{vassilakopoulou2021bridging}. This has helped to remove barriers to education and make education more equitable. Education 2.0 also started to address the shortcomings of Education 1.0 in terms of a one-size-fits-all approach, by incorporating blended learning, and distance learning, which allowed students to learn at their own pace, in their own time and place \cite{dakhi2020blended,traxler2018distance}. Furthermore, it started to provide opportunities for students to take an active role in their own learning and it fostered creativity and independent thinking, by encouraging the use of technology for collaborative and interactive learning \cite{jeong2019cognitive}.

However, despite its many advantages, it still had some shortcomings that led to the development of Education 3.0. One of the main shortcomings of Education 2.0 was that it mainly used technology as a supplement to traditional teaching methods, and it was not fully integrated into the one-size-fits-all approach of Education 1.0, and it still had a focus on the teacher as the primary source of information and students were still expected to be passive learners \cite{anggriani2020effectiveness}. Furthermore, it did not fully take advantage of the potential for collaborative and teamwork learning \cite{blau2020does}. Education 3.0 addresses these shortcomings by fully integrating technology into the teaching and learning process and emphasizing active and collaborative learning, personalization, and student-centered learning \cite{yurtseven2022flipped,dimitriadou2023critical}. It also introduced the concept of the “flipped classroom,” which allows for more personalized and student-centered learning \cite{xie2019trends}.

\subsection{Education 3.0}

Education 3.0 represents an evolution in the field of education, building upon the advancements of Education 2.0 and addressing its shortcomings. Education 3.0 fully integrates technology into the teaching and learning process and strongly emphasizes active and collaborative learning. One of the key features of Education 3.0 is the use of the “flipped classroom” approach, where students watch lectures and complete homework assignments at home and then use class time for discussion and interactive activities. This approach allows for more personalized and student-centered learning, allowing students to work at their own pace and on their own level. It also fosters critical thinking, creativity and problem-solving skills, which are important for the 21st century. \cite{al2020flipped}

Education 3.0 also strongly emphasizes collaboration and teamwork, encouraging students to take an active role in their learning. This allows students to learn from one another and to develop important 21st-century skills such as communication, collaboration, and critical thinking. \cite{van2019systematic} Additionally, Education 3.0 uses data and analytics to track student progress and identify learning gaps, which allows teachers to tailor instruction to meet the needs of individual students and make data-driven decisions to improve student outcomes. \cite{nguyen2020data}

Despite its many benefits over earlier Education paradigms, it still had some shortcomings that led to the development of Education 4.0. These shortcomings include the limited use of technology, which is primarily used for the delivery of content, limited opportunities for personalization and student-centered learning, lack of focus on well-being, stress, and mental health support for students and teachers, and limited opportunities for collaboration and teamwork. Education 4.0 addresses these shortcomings by fully integrating technology to support student-centered learning, collaboration, and personalization, placing a strong emphasis on well-being, stress, and mental health support for students and teachers, and utilizing advanced technologies such as Artificial Intelligence, Virtual and Augmented Reality, IoT, Cloud Computing, Big Data and Analytics, Blockchain and 5G networks to enhance the teaching and learning process. \cite{peng2019personalized} \cite{network2020impact} \cite{rani2021amalgamation} \cite{bhaskar2021blockchain}.

\subsection{Education 4.0}

This stage built on the concept of Education 3.0 and focused on using technology to enhance the learning experience. Education 4.0 is a holistic education approach incorporating the latest technologies to enhance the learning experience. It is built on the principles of personalized, student-centered, and adaptive learning and is designed to develop 21st-century skills such as digital literacy, collaboration, communication, and critical thinking. Recent research and technological breakthroughs have enabled the integration of AI and machine learning (ML) in education. AI and ML can be used to personalize the learning experience by analyzing student data and adapting the content and pace of instruction to match the student's individual needs and abilities \cite{vikhman2022technological,shawky2019towards,orji2020using,parmar2023open,shawky2018reinforcement}.

Virtual and augmented reality (VR/AR) are also being increasingly used in education to provide immersive and interactive learning experiences. VR/AR can be used to simulate real-world scenarios, allowing students to learn through hands-on experiences in a safe and controlled environment \cite{garzon2019meta,yuliono2018promising,egger2020augmented,al2019exploring,rojas2023systematic}. The IoT is also playing a significant role in Education 4.0, by enabling the integration of smart devices and sensors in the classroom. This allows for real-time monitoring and tracking of student engagement and progress and enables teachers to provide more targeted and personalized instruction \cite{hassan2023importance,razzaque2020internet,abichandani2022internet,lampropoulos2022augmented,sarker2022internet,sarker2022internet}.

In addition, the use of gamification in education is also on the rise, with games and interactive simulations being used to make learning more engaging and enjoyable. Gamification can also be used to teach complex and abstract concepts and to develop problem-solving and critical thinking skills. Consequently, it has significantly more positive changes and integration of technologies compared to another version, yet it faces some notable shortcomings, which led the researchers to start thinking about Education 5.0. For instance, One of the main shortcomings of Education 4.0 is the lack of accessibility, as it relies heavily on technology and the internet, which can create barriers for students who do not have access to these resources. Inequalities can also be exacerbated as the integration of technology and the internet in Education 4.0 can also exacerbate existing educational inequalities, with students from disadvantaged backgrounds being disadvantaged. Another shortcoming is limited human interaction, as the use of technology in Education 4.0 can lead to a decrease in human interaction, which is essential for developing social and emotional skills. Over-reliance on technology is also one of the shortcomings, as over-reliance on technology can create a situation where students become dependent on technology to learn and lack the necessary skills to learn independently.
Education 5.0 aims to address these shortcomings by creating a more inclusive and equitable education system that utilizes technology to enhance human interaction and promote independent learning. It will likely focus on more personalized and adaptive learning, with technology being used to enhance the human experience rather than replace it. Additionally, it will incorporate the latest advancements in technology, such as the use of big data, blockchain, and quantum computing, and it will focus on developing an interdisciplinary approach to learning, integrating different fields of study and fostering more collaboration between students and teachers.

\subsection{Summary and Insights}

In summary, the current stage, Education 5.0, represents a new level of technology integration in education. It builds on the concepts of the previous stages and incorporates advanced technologies such as Artificial Intelligence, Virtual and Augmented Reality, IoT, Cloud Computing, Big Data and Analytics, Blockchain and 5G networks to create a more efficient, effective, and equitable education system. Education 5.0 aims to promote personalized learning, collaboration, and well-being and to develop 21st-century skills such as critical thinking, creativity, and problem-solving by utilizing modern ICT and related technologies such as blockchain, IoT, AI, ML and AR/VR-based Metaverse. The evolution discussed in this chapter led to some of the crucial requirements for an Education 5.0 system, which is exhibited in the subsequent section.

\section{Requirements}

In order to realize the vision of Education 5.0, the core set of requirements is highlighted based on the current state of the art, as shown below.

\subsection{Personalized Learning}
Education 5.0 focuses on providing personalized learning experiences that adapt to the needs and abilities of individual students. This can be achieved using artificial intelligence and machine learning to create personalized learning plans and adjust instruction in real time based on student progress. Personalized Learning is an important aspect of Education 5.0, which aims to tailor the learning experience to each student's individual needs and abilities. Evaluation metrics used to measure the effectiveness of personalized learning may include student engagement, motivation, and achievement, as well as the ability of the instruction to adapt to the student's individual needs and abilities \cite{zhang2020understanding,raj2022systematic,zheng2022effectiveness}. Another important metric for evaluating personalized learning is the use of student data, such as formative and summative assessments, to inform instruction and track progress \cite{kye2021educational,zhong2023application,margot2019teachers}.

\subsection{Collaboration and Connectedness}
Education 5.0 promotes collaboration and connectedness among students, teachers, and other stakeholders. This can be achieved through the use of technology such as virtual and augmented reality and the IoT, which allow for immersive and interactive learning experiences. Evaluation metrics for collaboration and connectedness in Education 5.0 can include measures of student engagement and participation in group work and collaborative projects, as well as assessments of the quality and effectiveness of these collaborations. Additionally, communication and teamwork skills, such as conflict resolution and problem-solving abilities, can be used to evaluate collaboration and connectedness. The use of technology, such as social media and online collaboration tools, can also be evaluated for its effectiveness in promoting connectedness and collaboration among students and teachers \cite{herro2017co,mrklas2023tools,shi2023bridge}.

\subsection{Development of 21st-century skills}
Education 5.0 focuses on developing 21st-century skills such as critical thinking, creativity, and problem-solving rather than just rote learning. This can be achieved through the use of technology such as game-based learning, project-based learning, and other active learning methods. Evaluation metrics for developing 21st-century skills in Education 5.0 could include assessments of students' abilities in digital literacy, collaboration, communication, and critical thinking. These metrics include standardized tests, performance-based assessments, self-reflection, and peer evaluations. Additionally, observation of student engagement and participation in collaborative and technology-rich learning activities can also be used as an evaluation metric. Furthermore, incorporating project-based learning, problem-based learning, and other forms of experiential learning can also provide an opportunity to evaluate the development of 21st-century skills in students. \cite{ydesen2023assessment}

\subsection{Flexibility and Accessibility}
Education 5.0 aims to make education more flexible and accessible by removing barriers to education, such as geographic and financial constraints. This can be achieved through the use of technology such as cloud computing \cite{alam2022cloud}, which allows students to access digital resources and materials from anywhere and at any time. Flexibility and accessibility are two important characteristics of Education 5.0, which refers to the integration of technology in education to enhance the learning experience \cite{lantada2020engineering}.

Flexibility in education refers to the ability to adapt to different learning styles and needs \cite{lantada2020engineering}. A common metric used to evaluate the flexibility of an educational program is the Learner Satisfaction Index (LSI) \cite{wulandari2023complexity}, which measures student satisfaction with the program's flexibility. The LSI can be measured through surveys or interviews with students \cite{hai2022determinants}, and it can be used to identify areas of improvement in the program \cite{zhou2023analysis}.

Accessibility in education refers to the ability to provide education to all students, regardless of their physical, mental, or financial abilities. A commonly used metric to evaluate the accessibility of an educational program is the Universal Design for Learning (UDL) framework \cite{UDL_framework}, which measures the extent to which a program is designed to be accessible to all students. The UDL framework includes three main components: representation, action and expression, and engagement \cite{zhang2022integrating}. These components can be evaluated through the use of surveys \cite{radjabova2023improving}, interviews \cite{fornauf2020toward}, or observations \cite{ok2017universal}.

\subsection{Data-Driven Decision Making}
Education 5.0 relies on data-driven decision-making. Evaluating the success of Data-Driven Decision Making (DDDM) in Education 5.0 involves using a variety of metrics to assess student outcomes, learning analytics, technology adoption, user satisfaction, and return on investment. Student outcome metrics such as grades, test scores, graduation rates, and post-graduation success provide insight into student achievement \cite{caspari2023learning}. Learning analytics involves collecting and analyzing data about student learning, including engagement, interaction, and assessment data \cite{rivera2019exploring}. Technology adoption metrics evaluate the extent to which teachers and students are using educational technology \cite{kurilovas2020data}. User satisfaction metrics measure the satisfaction of students, teachers, and administrators with the DDDM process \cite{ganahl2023data}. Return on investment metrics assess the financial benefits of DDDM, such as cost savings and increased revenue \cite{taylor2020neoliberal}. \cite{botvin2023data}, using big data and analytics \cite{park2022big} to track student progress and identify learning gaps, and to tailor instruction to meet individual needs \cite{zheng2022effectiveness}.

\subsection{Security and Privacy}
Education 5.0 requires secure and private handling of student data, using blockchain technology to ensure data privacy and integrity. Ensuring the security and privacy of data in Education 5.0 is critical to successfully implementing this new educational model. The security of the data stored and transmitted in the educational system must be protected from unauthorized access, theft, and cyber attacks. The privacy of student data must be protected, including personal information such as name, address, and social security number, as well as academic data such as grades and test scores \cite{nair2023issues}.

Several metrics can be used to evaluate the security and privacy of the Education 5.0 system. These include data breach incidents, security and privacy audits, and user perceptions of data security and privacy \cite{vaiopoulou2022classification}. Data breach incidents can be evaluated by tracking the number of unauthorized accesses, data theft incidents, and cyber attacks on the system \cite{shaikh2023information}. Third-party security experts can perform security and privacy audits to evaluate the strength of the system's security and privacy measures \cite{nirmala2022efficacy}. User perceptions of data security and privacy can be evaluated through surveys or interviews with students, teachers, and administrators to determine their level of trust and confidence in the security and privacy of the system \cite{gogus2019privacy}.

\subsection{High-speed networks}
High-speed networks are an essential aspect of Education 5.0, as they facilitate real-time delivery of digital resources and materials to students and teachers. The use of high-speed networks in education enables seamless communication and collaboration between students, teachers, and educational institutions. Moreover, high-speed networks ensure that students can access educational materials remotely. To evaluate the performance of high-speed networks in education, several metrics can be used, such as network reliability, network capacity, and network latency.

\cite{haleem2022understanding} and \cite{williamson2019policy} have studied the impact of high-speed networks on the delivery of education services and found that high-speed networks have a significant positive impact on the quality of education.

\subsection{Well-being}
Education 5.0 strongly emphasizes student well-being, including physical, mental, and emotional health. This requires using technologies such as the IoT, which can monitor students' engagement and progress and provide teachers with real-time feedback on student performance. a variety of measures can be used to assess various aspects of well-being, such as mental and emotional health, physical health, and academic performance \cite{vaivada2022interventions}. Some common metrics include student satisfaction surveys \cite{ryan2022learning}, academic performance indicators (e.g., grades, test scores) \cite{sloane2003issues}, attendance and engagement metrics \cite{fredricks2019interventions}, mental health and wellness assessments \cite{russell2023assessing}, and physical health indicators (e.g. fitness test results, sport participation rates) \cite{demetriou2017after}. However, it's important to note that the specific metrics used may vary depending on the context and the individual goals of the educational institution \cite{mcgaghie2010critical}.

\subsection{Adaptability}
Education 5.0 is designed to be adaptable to the changing needs of the workforce and society. This requires the use of technologies such as cloud computing and blockchain, which allow for sharing of resources and materials and the secure storage of student data. Assessing the adaptability of Education 5.0 requires evaluating several key metrics. Flexibility, scalability, and interoperability are three important factors to consider. Flexibility refers to the education system's ability to adjust to changing needs and circumstances, such as workforce demand shifts or student population changes. Scalability refers to the capacity of the education system to expand or contract to meet changing needs and demands, such as growth in student enrollment or new educational programs. Interoperability refers to the ability of different education systems, institutions, and platforms to work together seamlessly and share data, resources, and materials. Another important aspect of adaptability is user-centeredness. This refers to the degree to which the education system is designed to meet the needs and preferences of students, educators, and other stakeholders. This can be evaluated through user satisfaction, engagement, and ease of use metrics.

In addition, data privacy and security are a critical components of adaptability in Education 5.0. As more data is generated and shared through educational systems, it is important to ensure that sensitive information is protected and not misused. Evaluation metrics for data privacy and security can include measures of data security, data privacy, and regulatory compliance. Finally, continuously improving and evolving is an important part of adaptability in Education 5.0. This can be evaluated through metrics such as the frequency and speed of updates, the effectiveness of new features and functions, and the degree to which the education system can incorporate feedback from users and stakeholders.

\subsection{Accessibility}
Education 5.0 aims to remove barriers to education and make education more accessible to all. This requires using technologies such as 5G networks and cloud computing, which can support the high-bandwidth requirements of many new educational technologies and make education more flexible and accessible. Assessing the accessibility of Education 5.0 is critical in evaluating its success. Accessibility refers to the degree to which educational resources and opportunities are available to individuals regardless of their location or socio-economic status. To evaluate the accessibility of Education 5.0 initiatives, it is important to consider a number of metrics. One metric to consider is the availability of educational resources, such as online courses, textbooks, and instructional materials. This includes not only the quantity of resources available but also their quality and relevance to the needs of students. A study by \cite{kirkwood2014technology} found that the availability and quality of educational resources is a critical factors in the adoption and effectiveness of technology-enhanced learning.

Another important metric is the availability and quality of internet connectivity, which is critical for accessing online resources and participating in virtual learning environments. Physical accessibility, including the availability of transportation and accessible facilities, must also be considered.  The demographic representation among individuals with access to educational resources, including gender, race, and socio-economic status, must also be examined. Finally, the cost of educational resources, including tuition and related expenses, can also be a barrier to access.

\subsection{Gamification and Game-based Learning}
Game-based learning is a crucial requirement for Education 5.0 because it can engage students in a way that traditional classroom teaching may not. Game-based learning can help increase motivation, engagement, and learning outcomes.   Evaluating the effectiveness of gamification in education is important to determine its impact on student learning outcomes. Several metrics can be used to assess the effectiveness of game-based learning. One important metric is student engagement, which refers to students' motivation, interest, and involvement in learning activities. A study by Papastergiou found that game-based learning can increase student engagement and motivation in computer science education \cite{papastergiou2009digital}. Another study by \cite{hosseini2019learning} showed that game-based learning can improve student engagement and motivation in science education, leading to higher academic achievement.

Another important metric is learning outcomes, which refers to the knowledge and skills that students acquire through game-based learning activities. A study by Cheng et al. found that game-based learning can improve learning outcomes in science education \cite{cheng2020does}. Another study by Deng et al. found that game-based learning was effective in teaching complex concepts in math education \cite{deng2020digital}.

Student feedback is also an important metric, as it provides insight into the students' perception of game-based learning activities and their effectiveness. A study by Hartt et al. found that student feedback was positive for game-based learning activities, indicating that students valued the interactive and engaging nature of these activities \cite{hartt2020game}.

Finally, the design and implementation of game-based learning activities must be considered, as they can impact the effectiveness of these activities. A study by Ramli et al. found that well-designed game-based learning activities can improve student engagement and learning outcomes \cite{ramli2020enhancing}.

\subsection{Summary and Insights}
The above requirements must be met to make an Education 5.0 system. We envision top five parameters for each requirement which can be used for the assessment of whether a particular requirement is met or not. Table \ref{tab:params} summarizes the requirements and the top 5 parameters for evaluation. In conclusion, these metrics provide valuable information for evaluating the effectiveness of gamification or game-based learning in Education 5.0. By considering these metrics, educators, and researchers can better understand the impact of game-based learning on student motivation, engagement, and learning outcomes.

\begin{tiny}
\begin{table}[h!]
    \centering
    \caption{Overview of various requirements of Education 5.0 and their top five evaluation metrics. 
     \footnotesize
     \label{tab:params}}
    \begin{tabular}{p{20pt}p{80pt}p{85pt}p{85pt}p{85pt}p{85pt}}
         \cline{1-6}
         \textbf{Label} & \textbf{M1} & \textbf{M2} & \textbf{M3} & \textbf{M4} & \textbf{M5}   \\ \cline{1-6}
         R1 &  Personalized Learning & Engagement & Motivation & instructors' adaptation ability & student data analysis \\ \cline{1-6}
         R2 & participation in groups & communication & teamwork assessment & use of collaboration tools \\ \cline{1-6}

         R3 & Digital literacy & critical thinking & self-reflection & problem-based learning & student engagement \\ \cline{1-6}

         R4 & LSI-based surveys & LSI-based interviews & UDL-based representation & UDL-based action and expression & UDL-based engagement \\ \cline{1-6}

         R5 & graduation rate & student satisfaction & student engagement & student achievement & technology adoption \\ \cline{1-6}

         R6 & data breach incidents & cyber attacks & student perception & level of trust & security and privacy audits \\ \cline{1-6}

         R7 & network reliability & network capacity & latency & jitter & ability of online streaming \\ \cline{1-6}

         R8 & student satisfaction & academic performance indicators & attendance & engagement & mental health \\ \cline{1-6}

         R9 & user-centeredness & data privacy & security & continuous improvement & evolution \\ \cline{1-6}

         R10 & Availability of resources & internet connectivity & physical accessibility & cost & demographic representation \\ \cline{1-6}

         R11 & student engagement & learning outcomes & student feedback & students motivation & involvement rate

         \\ \cline{1-6}
         
    \end{tabular}
    \\
    \footnotesize{\textit{\\R1: Personalized learning, R2: Collaboration and connectedness, R3: Development of 21$_{st}$-century skills, R4: Flexibility and accessibility, R5: Data-driven decision making, R6: Security and privacy, R7: high-speed networks, R8: Well-being, R9: Adaptibility, R10: Accessibility, R11: Gamification and Game-based learning.}}

\end{table}

\end{tiny}

\section{Enabling Technologies}
In this section, we will elucidate the enabling technologies that will act as a building blocks for Eduction 5.0. We envision that the enabling technologies for Education 5.0 include but are not limited to Artificial Intelligence (AI),Virtual and Augmented Reality (VR and AR), Internet of Things (IoT), Big Data and Analytics, Blockchain and 5G Networks. While AI-powered tools can personalize learning experiences, provide real-time feedback, and assist teachers in assessing students' progress, VR and AR can provide immersive and interactive learning experiences, allowing students to explore new environments and subjects in a more engaging way.

\begin{figure}[h!]
\centering
\includegraphics[width=\textwidth]{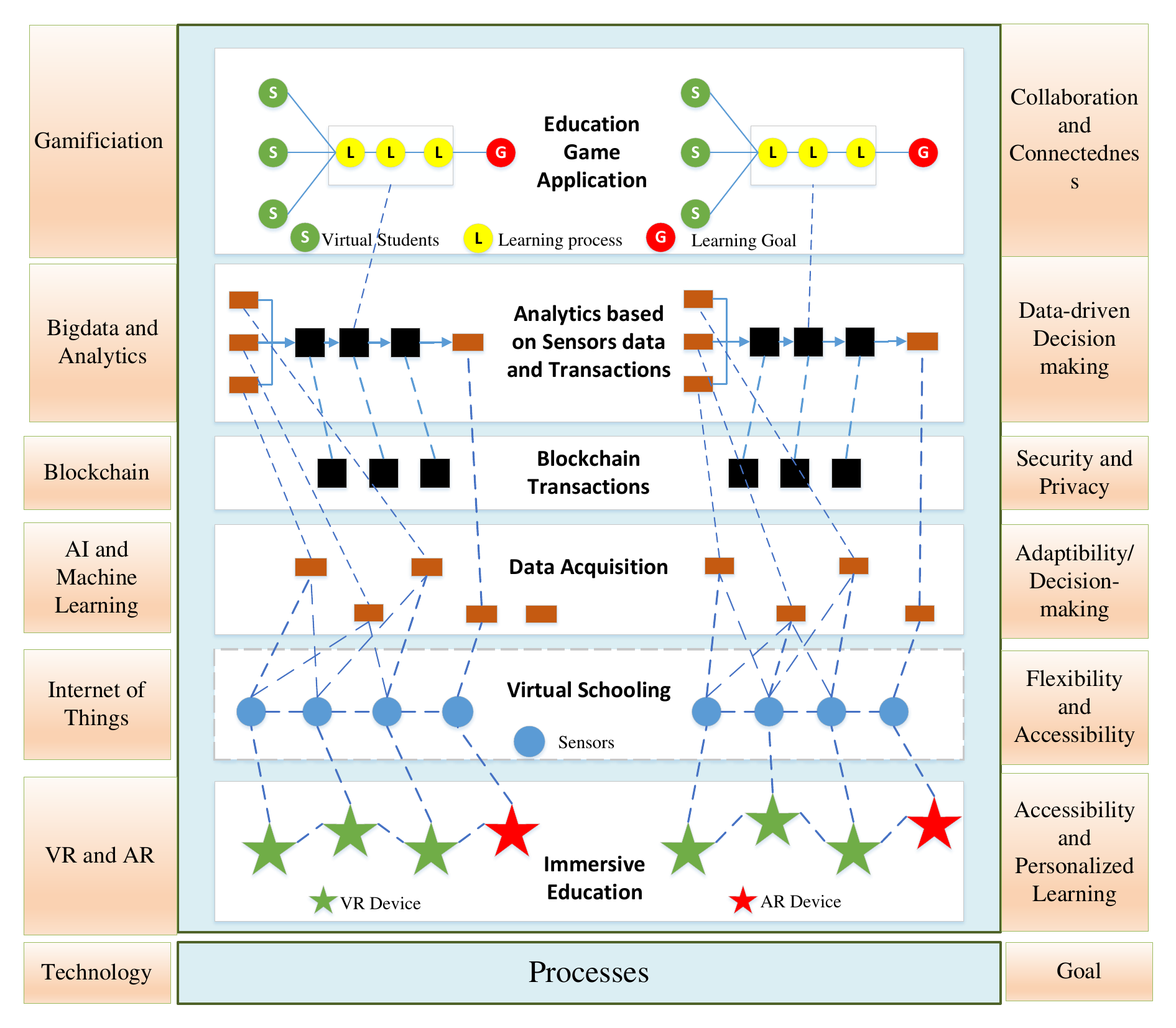}
\caption{\label{fig:tech} Two-way stack of Education 5.0. The left side depicts the technologies while the right side represents the goals to achieve. For instance, the use of AI and machine learning assess adaptive learning and decision-making. The central part exhibits the processes which form the overall learning methodology.}
\end{figure}

Additionally, IoT can be used to connect various devices and sensors to the internet, allowing for real-time monitoring and analysis of student engagement and progress wheras Cloud computing enables access to digital resources and materials from anywhere and at any time, making education more flexible and accessible. Moreover, Big data and analytics can be used to track and analyze student progress, identify learning gaps, and tailor instruction to meet individual needs whereas Blockchain can be used to secure and share student data, and ensure data privacy and integrity. Finally, 5G Networks can support the high-bandwidth requirements of many new educational technologies, such as virtual reality, and help to provide a seamless learning experience. In the following subsections, we will outline the progress being made in these core enabling technologies towards the realization of the vision of Education 5.0. The overall 2-way representation of Education 5.0 end its enabling technologies are portrayed in Fig \ref{fig:tech}.

\subsection{Artificial Intelligence}
Computers, machines, and other artifacts now exhibit human-like intelligence that is defined by cognitive capacities, learning, adaptability, and decision-making capabilities thanks to the field of research known as artificial intelligence (AI) and the inventions and developments that have followed. According to the study, AI has been widely adopted and employed in education in a variety of ways, especially by educational institutions \cite{celik2023towards}. Computers and computer-related technologies were the first forms of AI, which later evolved into web-based and online intelligent education systems, embedded computer systems, and other technologies \cite{arun2023review}. The use of humanoid robots and online chat-bots to carry out the tasks and obligations of instructors alone or in conjunction with teachers is an excellent example. These platforms have helped teachers improve the quality of their instructional activities and carry out other administrative tasks, such as reviewing and grading students' assignments, more quickly and effectively \cite{alam2022employing}. The systems make use of machine learning and flexibility, and the curriculum and content have been customized to meet the needs of the students. This has encouraged uptake and retention, which has enhanced the learning experience for students as a whole\cite{schiff2022education}. All researchers in the field of artificial intelligence in education (AIED) are virtually undoubtedly driven by moral considerations, such as enhancing students' possibilities for lifelong learning \cite{xia2022self}.

The fact that adaptive AIED systems frequently collect a lot of data, which can then be computed to dynamically improve the pedagogy and domain models, is one of their advantages. This method not only tests and improves our comprehension of the processes of teaching and learning, but also provides new information on how to offer support that is more effective, individualized, and contextualized \cite{huang2023effects}. 
We concentrate on two groups of AIED software programs: intelligent support for collaborative learning and personal tutors for every learner. These programs have been built to directly help learning.

\subsubsection{Personalized Learning based on AI}

The tendency toward lifelong learning and self-directed learning has been pushed by the pandemic, intelligent tutors have developed into a crucial tool for fostering both individual learning and teacher-student interaction \cite{chen2022two}. Intelligent tutor applications use AI to create a personalized learning path for each student depending on their accomplishments and needs and to help students understand a specific topic or lesson \cite{maghsudi2021personalized}. AI examines learning patterns to assist students in learning more effectively and acquiring the necessary skills. Intelligent or personalized studies are focused on tackling the largest barriers to education\cite{nguyen2022self}. The individualized learning experience, which is accessible anywhere and at any time, has become the norm. The core tenet of personalized learning is the automation of the teaching process and the function of the human tutor\cite{bhutoria2022personalized}. A tutor is a private educator who offers one-on-one instruction; this kind of teaching has been found to be more successful than the conventional approach of one teacher per classroom \cite{raj2022systematic}. Unfortunately, not everyone is able to afford a private tutor, but an intelligent AI-based tutor is less expensive and affordable. A personalized study program powered by Intelligent AI not only fills in knowledge gaps for learners but also gives parents the tools they need to better assist their kids' learning. Intelligent AI-based personalized study also helps the optimization of learning ways and decision tree progress \cite{iyer2022advancing}. Current AIED, in contrast, don't have pre-determined learning paths, which means optimizing instructional methods in accordance with the performance of each student \cite{peng2022ai}. 

\begin{figure}
\centering
\includegraphics[width=0.6\textwidth]{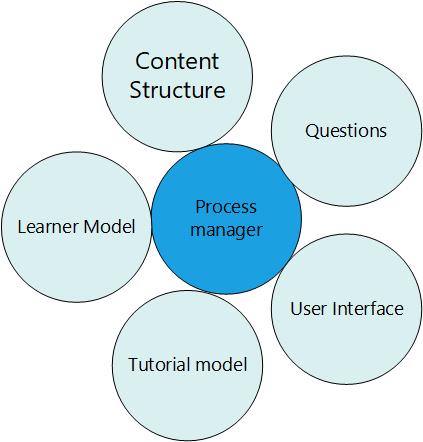}
\caption{\label{fig:AI_In_ED}Illustration of Personalized Learning based on AI.}
\end{figure}

To better meet each student's unique learning needs, AIED replicates personal human tutoring, offers focused real-time feedback, and creates tailored learning ways. AIED adjust to each learner and encourage them to focus on their own challenge areas as users gain more knowledge since they are more aware of what additional support each learner needs to progress. AIED is able to make the learning process personalized, engaging, inclusive, and flexible leading to improved study results.

\subsubsection{Collaborative Learning based on AI} 

Collaborative learning based on AI refers to the use of artificial intelligence and machine learning techniques to enhance the process of learning in a collaborative setting \cite{xu2022systematic}. This approach leverages technology to support communication, collaboration, and knowledge sharing between individuals and groups, with the goal of improving the learning experience and outcomes \cite{celik2023towards}. This can include AI-powered tools such as natural language processing, recommendation systems, and adaptive learning algorithms that can support students in their learning process by providing personalized feedback and guidance \cite{alhazmi2023ai}.
AI techniques can make educational settings accessible to all students, including those who have hearing or vision impairments \cite{wang2023artificial} or speak various languages \cite{malakul2023effects}, on a worldwide available.
In this approach, AI-powered tools and algorithms are used to facilitate communication, collaboration, and knowledge sharing between students and teachers \cite{southworth2023developing}. This can include features such as personalization, real-time feedback, and adaptive learning, which can help students learn more effectively and efficiently. Additionally, AI-based education collaborative learning can also support educators in their teaching, by providing insights into student performance and facilitating data-driven decision-making. The goal of this approach is to improve the overall learning experience and outcomes for students.
\subsection{Internet of Things}

Education 5.0 is the next evolution in education, which aims to leverage modern technologies to achieve the goal of lifelong learning. One of the key technologies that plays a pivotal role in Education 5.0 is the Internet of Things (IoT) \cite{hassan2023importance}; \cite{gul2017survey}. IoT-enabled devices such as wearables, sensors, and smart devices can be used to collect real-time data on student learning and behavior, providing a deeper understanding of individual student needs and abilities \cite{kassab2020systematic}; \cite{gul2017survey}.

IoT devices can also be used to facilitate personalized learning, an approach that leverages technology to individualize instruction, by providing adaptive learning systems that adjust the pace and content of instruction based on student performance \cite{hassan2023importance}; \cite{gul2017survey}. Personalized feedback can also be provided to students through IoT devices, such as wearables that track movements and provide feedback on posture and movement patterns \cite{gul2017survey}; \cite{kassab2020systematic}. Smart content, which adapts to the individual needs and abilities of each student, can also be created using IoT devices, such as connected textbooks with interactive elements and multimedia resources.

IoT devices can also be used to facilitate collaborative learning, an approach that emphasizes group work and the sharing of knowledge and skills among students, by enabling students to connect and collaborate with each other and with educational resources from anywhere, creating a more dynamic, interactive, and engaging learning environment for students.

Furthermore, IoT devices can be used to create smart classrooms that facilitate collaboration among students and teachers. This can enhance their collaborative skills and improve their learning outcomes.

Overall, the use of IoT devices in education can facilitate the implementation of personalized and collaborative learning, resulting in a more effective and efficient learning experience for students.

\subsubsection{Personalized Learning}
Personalized learning, an approach that leverages technology to individualize instruction, has gained significant attention in recent years \cite{ling2022use}; \cite{siemens2011penetrating}. IoT devices, such as wearables and sensors, have emerged as powerful tools for implementing personalized learning in the classroom \cite{bhutoria2022personalized}; \cite{raj2022systematic}. By collecting real-time data on student learning and behavior, IoT devices can provide a deeper understanding of individual student needs and abilities \cite{raj2022systematic}; \cite{zeeshan2022internet}. Furthermore, adaptive learning systems can be implemented using IoT technology, allowing for adjustments in pace and content of instruction based on student performance. Personalized feedback can also be provided to students through IoT devices, such as wearables that track movements and provide feedback on posture and movement patterns. Smart content, which adapts to the individual needs and abilities of each student, can also be created using IoT devices, such as connected textbooks with interactive elements and multimedia resources  \cite{siemens2011penetrating}. Lastly, IoT devices can be used to monitor student engagement and behavior in real time, providing teachers with valuable insights to improve their instruction. Overall, the use of IoT devices in education can facilitate the implementation of personalized learning, resulting in a more effective and efficient learning experience for students \cite{zeeshan2022internet}; \cite{siemens2011penetrating}.

\subsubsection{Collaborative Learning}
Collaborative learning, an approach that emphasizes group work and the sharing of knowledge and skills among students, has been shown to be an effective way to improve student learning outcomes \cite{dillenbourg1999collaborative}; \cite{johnson_johnson_1991}; \cite{wang2021evolution}. The use of Internet of Things (IoT) devices in education can facilitate collaborative learning by enabling students to connect with each other and with educational resources from anywhere \cite{wang2021evolution}; \cite{strauss2020promoting}.

One of the key aspects of collaborative learning using IoT is the ability to facilitate remote and distributed collaboration among students, regardless of their location \cite{strauss2020promoting}; \cite{hernandez2019computer}. IoT-enabled devices such as smartphones, tablets, and laptops can be used to share documents, communicate and collaborate on projects in real-time \cite{hernandez2019computer}. Additionally, virtual and augmented reality (VR/AR) can be used to create immersive virtual and augmented reality experiences that enable students to collaborate and interact with each other in a shared digital space \cite{wang2021evolution}; \cite{hernandez2019computer}.

Furthermore, IoT-enabled devices can be used to create smart classrooms that facilitate collaboration among students and teachers. For example, smartboards can be used to display and share information, and students can use connected tablets to participate in interactive activities and quizzes.

\subsubsection{Remote Learning}
Remote learning, also known as distance learning or e-learning, refers to the ability to access educational content and resources from a remote location, typically through the use of technology such as computers, smartphones, or tablets. With the advent of the Internet of Things (IoT), remote learning has become an increasingly important aspect of Education 5.0.

IoT technology can facilitate remote learning by enabling students to access educational content and resources from anywhere, at any time, using a variety of devices such as smartphones, tablets, and laptops \cite{gupta2023teachers}, \cite{lin2022metaverse}. This allows students to learn at their own pace, and to access educational resources that may not be available in their immediate environment.

IoT-enabled devices such as wearables and sensors can also be used to collect real-time data on student learning and behavior, providing a deeper understanding of individual student needs and abilities  \cite{lin2022metaverse}. This information can be used to create personalized learning experiences that are tailored to the unique needs and abilities of each student.

Furthermore, IoT technology can be used to facilitate remote collaboration among students and teachers, regardless of their location. For example, students can use IoT-enabled devices to share documents, communicate, and collaborate on projects in real-time, while teachers can use IoT-enabled devices to monitor student engagement and provide feedback on student progress.

Virtual and augmented reality (VR/AR) can also be used in remote learning to create immersive virtual and augmented reality experiences that enable students to collaborate and interact with each other in a shared digital space \cite{gupta2023teachers}, \cite{rojas2023systematic}. This can enhance their collaborative skills and improve their learning outcomes.

Overall, the use of IoT technology in remote learning can facilitate the implementation of personalized and collaborative learning, resulting in a more effective and efficient learning experience for students.

\subsection{AR, VR, MR, XR, and Metaverse}
Augmented Reality (AR) \cite{lampropoulos2022augmented, saundarajan2020learning}, Virtual Reality (VR) \cite{alam2022employing, gupta2023teachers, rojas2023systematic}, Mixed Reality (MR) \cite{leng2022industry, MR2}, and Extended Reality (XR) \cite{leng2022industry, XR1} are different forms of technology used to create immersive experiences. AR enhances the real world by overlaying digital information on the physical environment. This allows users to see the real world while looking at virtual objects, creating a hybrid environment. On the other hand, VR creates fully simulated environments, allowing users to experience and interact with computer-generated worlds as if they were real. MR is a combination of AR and VR, allowing virtual objects to interact with the physical environment. Meanwhile, Metaverse \cite{leng2022industry, MR2} is a virtual shared space created by the fusion of the physical and virtual worlds, where users can interact with each other and interact with digital objects in a seemingly real environment. XR is an umbrella term that encompasses AR, VR, MR, and other related technologies used to create immersive experiences.

Education 5.0 is a revolutionary approach to education, leveraging the latest technology to provide students with a personalized and adaptive learning experience. The usage of AR, VR, MR, and XR is transforming the way students learn, making education more interactive and engaging. The use of these technologies is explained the in the following.

Augmented Reality (AR) technology allows users to layer virtual objects onto their physical environment, making learning more interactive and visually engaging. With Education 5.0, teachers can use AR to provide students with interactive learning materials such as animations, videos, and 3D models. This technology helps bring textbooks to life and makes learning more engaging. For example, students can use AR to view virtual images of the human body and study anatomy by examining different parts in detail.

Virtual Reality (VR) technology creates fully simulated environments that allow users to experience and realistically interact with computer-generated worlds. Education 5.0 lets users use VR to simulate real-world scenarios and deliver hands-on learning experiences without being physically there. This technology provides students with a more immersive and interactive learning experience and helps make teaching more engaging and effective. For example, students can use VR to experience historical events, participate in virtual field trips, and practice skills in simulated environments.

Mixed Reality (MR) technology combines the real and virtual worlds, allowing virtual objects to interact with the physical environment. With Education 5.0, MR can be used to deliver blended learning experiences, allowing students to enhance physical textbooks with digital information and animations. This technology can help make learning more interactive and engaging, and can also provide students with more information and resources than can be accessed in traditional textbooks. For example, students can use MR to explore virtual models of historic sites and landmarks and manipulate virtual objects in real-time.

Extended Reality (XR) technology is an umbrella term that encompasses AR, VR, MR, and other related technologies used to create immersive experiences. Education 5.0 can use XR to deliver seamless, integrated learning experiences that bridge the real and virtual worlds. This technology will help create a new age of education that is personalized, adaptive, and focused on student success.

Finally, Metaverse is a virtual shared space created by the fusion of the physical and virtual worlds, where users can interact with each other and interact with digital objects in a seemingly real environment. In Education 5.0, teachers can use the Metaverse to provide students with a virtual learning environment where they can participate in virtual classes, interact with classmates and teachers, and participate in virtual excursions. Not only can this technology help make education more accessible, allowing students to participate in classes anytime and anywhere, but it can also provide students with a more immersive and interactive learning experience.

In conclusion, the use of AR, VR, MR, XR, and Metaverse technologies in Education 5.0 has the potential to revolutionize the way students learn, making education more accessible, engaging, and effective. These technologies provide students with personalized and adaptive learning experiences, making teaching more interactive, immersive, and effective.

\subsection{Blockchain}

Blockchain technology has been identified as a key enabler of Education 5.0 due to its potential to transform the way educational data is managed and used \cite{arndt2019overview}. The decentralized and secure nature of blockchain allows for a secure and tamper-proof way of storing and sharing educational data, such as student records, credentials, and educational achievements. This helps to improve data privacy and security, and reduces the risk of data breaches and other security threats \cite{arndt2019overview}.

In the area of verification and credentialing, blockchain can provide a secure and reliable way of tracking and verifying educational achievements through the use of digital credentials that are immutable and easily verifiable \cite{savina2019trends}. This not only helps to ensure the authenticity of educational credentials but also provides a streamlined way for individuals to demonstrate their qualifications and skills to potential employers or further educational institutions.

Furthermore, blockchain technology has the potential to increase access to education by enabling the creation of decentralized educational platforms \cite{hjalmarsson2018blockchain}. These platforms can provide accessible, affordable, and equitable education for all, regardless of their location or financial background. This can help to address the issue of unequal access to education and provide opportunities for individuals to learn and grow.

Another important aspect of blockchain technology is its ability to increase transparency and accountability in the education sector \cite{park2021promises}. This can help to build trust between educators, students, and other stakeholders by providing a clear understanding of the educational landscape. This increased transparency and trust can also help to improve the overall quality of education by enabling stakeholders to make informed decisions based on accurate and up-to-date information.

\subsection{Big Data Analytics}

Big data analytics refers to the use of advanced data processing and analysis techniques to extract valuable insights from large and complex data sets. Educational process information analysis and decision-making processes in areas including academic success, faculty effectiveness, organizational expansion, and technical efficiency are being transformed by big data \cite{khaw2023influence}.

At the moment, AIEd uses learning analytic and data mining to provide social, cognitive, or emotional views to comprehend the cognitive changes in collaborative learning \cite{caspari2023learning}.By analyzing data from various sources, such as student test scores, attendance records, and demographic information, educators can gain insights into what factors contribute to student success and identify areas where they need to focus their efforts to improve student outcomes\cite{klavsnja2017data}. Big data analysis can help teachers understand each student's strengths, weaknesses, and learning preferences, allowing them to provide customized educational experiences and improve student engagement to improve personalizing learning \cite{attaran2018opportunities}. Recently big data analytics  using to enhance research and gain insights into learning patterns and educational outcomes, which can inform the development of new teaching methods and educational technologies. To improve administrative processes big data analytics streamline administrative processes, such as student enrollment, scheduling, and data management \cite{aseeri2023organisational}. Porter's value chain model is described by Ghazwan H \cite{hassna2022big} as a useful tool for pinpointing chances for big data analisys to enhance the higher education value chain and explained how big data analitics enhance the production of value in higher education. 
Using a partial data analysis model, Wanli X et al. \cite{xing2022understanding} presented research using the social cognitive theory to pinpoint potential environmental, individual, and behavioral elements that influence students' usage of data. In order to extract education behavior from large-scale student open-ended answers and to verify the convergent validity of the results by comparing them with theory-driven, Bilge G. et al \cite{gencoglu2023machine} provide a latent Dirichlet allocation topic modeling study along with a visualization tool.
\begin{figure}[h!]
\centering
\includegraphics[width=0.6\textwidth]{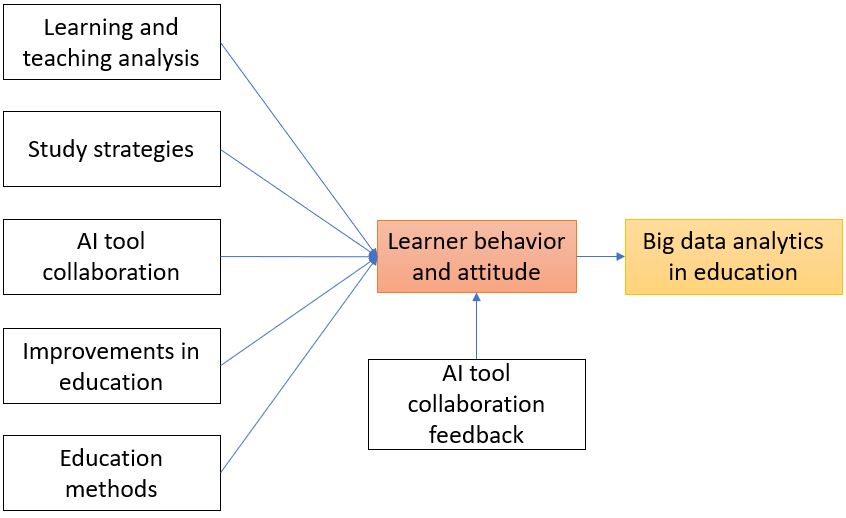}
\caption{\label{fig:Big_data}Illustration of Big data analytics in education system.}
\end{figure}
\subsection{Gamification}

Game-based learning is designed to make learning more interactive and engaging by incorporating elements of games such as points, leaderboards, and rewards. This can help increase student motivation and engagement, as well as improve learning outcomes. Additionally, game-based learning can promote problem-solving and critical thinking skills, which are important 21st-century skills. Furthermore, game-based learning can be personalized to the needs of individual students, this allows for students to work at their own pace and on their own level, which can help improve learning outcomes for all students, regardless of their ability level. Moreover, game-based learning can be used to create immersive and interactive learning experiences that can help students visualize and understand difficult concepts, this can be especially effective in STEM subjects.

There are several research papers that have studied the effectiveness of game-based learning in education.

For example, a study published in the Journal of Computer Assisted Learning found that game-based learning can improve student motivation, engagement, and learning outcomes in science education. The study showed that students who used game-based learning in a science classroom had a better understanding of the subject matter and were more motivated to learn.

Another study published in the Journal of Educational Technology Development and Exchange found that game-based learning can help improve problem-solving and critical-thinking skills in students. The study showed that students who used game-based learning had better problem-solving and critical thinking skills compared to students who did not use game-based learning.

Research conducted by the International Journal of Emerging Technologies in Learning found that game-based learning can create immersive and interactive learning experiences that can help students visualize and understand difficult concepts. The study showed that students who used game-based learning in a mathematics classroom had a better understanding of the subject matter and were more motivated to learn.

\subsection{Insight and Lesson Learned} 

Gamified learning of game-based learning is a crucial requirement for Education 5.0 because it can help increase student motivation, engagement, and learning outcomes, promote problem-solving and critical thinking skills and create immersive and interactive learning experiences that can help students visualize and understand difficult concepts. In conclusion, existing research has shown that game-based learning can improve student motivation, engagement, and learning outcomes, promote problem-solving and critical thinking skills, and create immersive and interactive learning experiences that can help students visualize and understand difficult concepts. These findings support the idea that game-based learning is a crucial requirement for Education 5.0.

Moreover, the recent shift towards Generative AI like ChatGPT and similar chatbots can provide instant, personalized support to students. Using AI and natural language processing, ChatBots can quickly answer questions, provide feedback on assignments, and even suggest resources to help students better understand the material. This can help students who are struggling to keep up with the pace of the course or who have special needs outside of class. Additionally, Chatbots can help create more engaging and interactive learning experiences. For example, a ChatBot can be a virtual tutor or mentor, guiding students through interactive learning activities and providing feedback on their progress. This makes learning more fun and engaging for students and allows for more individualized attention. 

Finally, the metaverse could also significantly impact on the way current education systems are operating. For instance, injecting VR/AR into the classroom has improved the learning experience and make it more interesting and playful in contrast to the traditional teaching methods.

\section{Current state-of-the-art and Future roadmap}
The existing benchmarks for AI in Education 5.0 are described from three perspectives. The first perspective is centered on students, the second on instructors, and the third on institutions.
\subsection{Student-focused AI in education 5.0 }
A small digression is required before analyzing the different benchmarks and modern tools of student-focused AI in education, i.e., AI-assisted tools expressly developed to aid students. Not all AI-assisted technology utilized by students have been intended for students. Instead, these technologies have been "repurposed" for educational purposes. Typically, these technologies are not considered purely to assist students, but they must be accounted for in any exhaustive overview of student-focused AI in education. Google's suite of collaborative tools, which includes Google Docs \cite{zioga2020collaborative}, is probably the most powerful AI-assisted technology that has been repurposed for education. In addition, social networking services such as WhatsApp \cite{cetinkaya2017impact}, ZOOM \cite{stefanile2020transition}, and content-sharing platforms such as YouTube \cite{margallo2023characterizing} are increasingly being utilized to promote student learning in a variety of ways (a trend that escalated during recent pandemic lockdowns across the globe). The following benchmarks for student-focused AI in education are described in detail. 
\subsubsection{ Smart teaching systems }
Smart teaching systems  (STS) that are increasingly advanced are the most widespread and undoubtedly best-funded AI deployments in education. Typically, they deliver computer-based, step-by-step instruction across topics in clearly delineated, organized courses. An STS provides a variety of individualized instruction, tasks, and assessments to each learner. While the learner participates in a certain activity, the system collects many points, such as what is viewed, what is written, which objectives have been successfully completed, and any misunderstandings that have been exhibited. This data is analyzed based on artificial intelligence training models  to identify the next information, activity, and quiz to be supplied, thus creating a personalized track through the content to be taught. The process is then repeated.STS contains a teacher dashboard that helps to assess student outcomes accurately. As an example iTalk2Learn \cite{grawemeyer2017affective} is a project developed in Europe; the developers of the system require that machine learning is used to compile personal lesson plans. 
\subsubsection{ AI based Applications }
There is a rapidly expanding collection of commercially accessible AI-assisted educational applications in the top online services. There are, for instance, increasingly amazing AI-assisted mathematics applications, such as Photomath \cite{saundarajan2020learning}, which some predict could make mathematics education to a new horizons. These apprehensions resemble those that accompanied the introduction of calculators in schools roughly five decades ago: if the tool can do it (automatically calculate a basic arithmetic, automatically translate between languages, or instantly determine solutions), there may be no need for children to learn how to do it, thereby impacting learning. Prahani \cite{prahani2022artificial} creates artificial intelligence-based solutions for K-12 and higher education organizations. Additionally, it is used in corporate training situations. One of Prahani's most important AI tools is its virtual learning assistant, which uses conversational technology to aid students in formulating open-format replies and enhancing their critical thinking abilities. In addition, the virtual assistant delivers tailored one-on-one teaching and real-time feedback for each student.
\subsubsection{ VR/AR/Mr based learning }
Virtual Reality (VR) and Augmented Reality (AR) simulation models and virtual games-based learning are regularly integrated with artificial intelligence machine learning, computer vision, and natural language processing and are increasingly employed in educational contexts. The advancement of AI-assisted display technologies, virtual reality (VR), augmented reality (AR), and mixed reality (MR) have been extensively used in thoracic surgery to convert 2D images into 3D models, which aids surgical education, planning, and simulation \cite{mumtaz2022future}. Google has created over a hundred VR and AR Expeditions appropriate for academic contexts. Likewise, digital games-based learning (DGBL) is progressively using artificial intelligence (AI) technology to personalize gaming to each learner \cite{zhan2022systematic}.
\subsubsection{AI-based Education 5.0 for People with Disabilities}
Education 5.0 is concentrated on the application of AI techniques for the identification of learning disorders such as ADHD (e.g., \cite{anuradha2010diagnosis}), dyslexia \cite{nilsson2016screening}, and dysgraphia \cite{asselborn2020extending}.In addition, a substantial study has been conducted on using robots in education 5.0 , particularly to assist youngsters on the autistic spectrum \cite{alabdulkareem2022systematic}. Several predominant AI technologies, such as text-to-speech applications and automated image tagging, have been reconfigured for kids with learning disabilities, along with a limited number of specialized AI-assisted apps, such as those that automatically interpret for children with hearing impairments.
\subsubsection{Advance Generative Pre-trained Transformers for learners }
The introduction of Chat GPT \cite{firat2023chat} can be mile stone towards education 5.0. Globally, essays continue to be a vital part of educational evaluation, however, plagiarism has been a widespread practice for centuries. With online essay mills selling customized writings on any subject, the Internet has facilitated this process with chat GPT alike tools. Recent AI advancements known as 'big language models,' such as the GPT-3 \cite{dehouche2021plagiarism} from Open AI outlined above, are set to have an even larger effect (GPT-3, 2020). There are currently a number of commercial companies that provide students with automated essay writing (AEW) systems that, in response to a stimulus such as an essay question, may write single paragraphs or full articles. Numerous companies now provide students with automated essay writing (AEW) programs that can automatically write individual paragraphs or full compositions in response to a topic such as an essay question. Despite the fact that the writing created by AEW is sometimes shallow and illogical, it is often impossible to discern whether the content was generated by a chatbot or a human. It is uncertain if AEW tools promote or hinder student learning. Nevertheless, because of their rising complexity and what might be defined as a rivalry among AEWs and AEW analyzers, they are likely to have an effect on how we evaluate learners \cite{selwyn2019should}.
\subsubsection{Automated Formative Assessment Model for Learning}
Automated formative assessment in education 5.0  employs natural language processing, semantic processing, and other AI-assisted approaches to deliver meaningful feedback on learner outputs. There are fewer researched and commercial automated formative assessment programs that exist despite their promise to promote student learning; this is likely due to the challenges associated with automatically giving accurate and useful feedback \cite{foster2019barriers}. Principally, no AI algorithm is now capable of the level of comprehension or accuracy of assessment that an instructor can provide, rather depending on the superficial characteristics of the writing. Research conducted at Stanford University examined an auto grader automated formative assessment system that offered feedback on coding assignments by 12,000 computer science students. Approximately 98 percent of programmers agreed with the provided comments, somewhat higher than their acceptance of instruction from human teachers \cite{metz2021making}.
\subsection{Teachers focused Eduction 5.0}
Many student-centered education, particularly STS, provide portals or dashboards for instructors, often based on open learner models, that provide a dynamic depiction of what individual learners and groups of learners have accomplished or their mistakes \cite{bodily2017review}.One innovative method involves the use of augmented reality (AR) spectacles used by the instructor to overlay dashboard-like statistics over the heads of their learners as the students participate in an STS \cite{holstein2021equity}.Although amazing, this is an example of employing an AI technology to solve a problem produced by an AI technology (in this case, to address the fact that when students are using an STS, their instructor cannot readily see what they are doing and hence cannot give proper assistance).Regardless, STS and other dashboard-enabled AI in education 5.0 are primarily designed with the student in mind. In fact, if we disregard overlaps for the sake of study, there are few instances of truly teacher-focused AI in education 5.0. Here, we consider six contentious prospects: plagiarism identification, intelligent curation of learning resources, classroom supervision, automated summative evaluation, AI teaching assistants, and classroom coordination.
\subsubsection{Detection of plagiarism}
Educators make extensive use of commercially accessible plagiarism detection technologies, for which machine learning techniques have been extensively adopted in the recent year. The teachers and institutions widely uses Turnitin \cite{mphahlele2019use} an AI driven plagiarism detector. Education 5.0 greatly focuses on  detection as there are widely available GPTs which are trained on internet  data proned to hugely plagiarised data.   
\subsubsection{Pedagogical supervision}
AI-assisted pedagogical supervision technologies that have been researched, developed, and made commercially accessible are becoming more prevalent.For instance, AI-assisted video apps have been designed to monitor where a student is looking, allowing the system to determine whether or not the student is concentrating on the instructor or the job at hand \cite{muzammul2019education}.Students are being requested to wear portable EEG (electroencephalography) headsets to monitor their cognitive function, which is potentially even more invasive.Across many academic institutions, AI-assisted technologies are also used to track a student's campus activities (often through a mobile app), what they obtain from the online learning system, and what they purchase at campus book shops and cafeteria \cite{alam2022employing}.
\subsubsection{AI based smart grading}
There has been anticipation for a long time that AI would help instructors save time and increase productivity the tedious and expensive task of grading student projects, tests, and other tasks\cite{watters2023teaching}. AI based auto-graders have been introduced for the assessment of student written tasks \cite{ramesh2022automated}.Some state-of-the-art autograders additionally conclude to identify the nature of the issue and propose to the learner how to remedy it, while others, dependent on the discipline, offer to evaluate student responses with approximately 90 percent reliability \cite{hsu2021attitudes}. 
\subsubsection {Technology enhanced classroom orchestration}
Orchestration is an approach to Technology Enhanced Learning that focuses on the issues of integrating technology into the classroom, with a special emphasis on strengthening instructors' responsibilities.While still in its infancy, research into the ways in which AI may aid classroom orchestration is expanding \cite{song2021review}.To assist the instructor in orchestrating such a complicated curriculum design, we designed a tablet application that enabled the teacher to observe the real-time status of the class, regulate the flow of activities, and know when and where he was required within the flow of class activities. The tablet used a collection of specifically created real-time software agents to handle student interactions in real time, allowing for the dynamic coordination of student groups, material distribution, and instructor alerts \cite{tissenbaum2019supporting}.
\subsection{Institution focused education 5.0}
Institutional priorities in education 5.0 include technology that facilitate the distribution of financial assistance, curriculum, scheduling, and work scheduling, and the identification of dropouts and students at risk, as well as the quality of education and learning \cite{kitto2020towards}.These technologies have a distinct administrative purpose and have many similarities with business-oriented Artificial Intelligence. In light of this, we will only discuss two essential and contentious institution-focused education topics: admissions (one of the high-risk use cases outlined in the proposed EU AI Act) and web security for educational institutions. 
\subsubsection{Admissions}
Not without controversy, many institutions of higher education, mostly in the United States, employ commercially accessible AI-assisted admissions software to help their admissions procedures. The objective is to minimize expenses and improve the fairness of the admissions process by removing invisible human biases (such as groupthink and racial and gender prejudices) that might influence decision-making \cite{holmes2022state}.
\subsubsection{Web security for institutions}
Early on in the COVID-19 epidemic, a substantial amount of instruction and assessments went online, resulting in the explosive growth of various exam-monitoring and web-security organizations \cite{nigam2021systematic}.
Web security strives to ensure academic integrity by employing AI-assisted cameras and microphones to automatically monitor students taking an online test by scanning their faces and recording their keystrokes and mouse movements \cite{chen2013security}.They are suspected of interference, incompetence, prejudice, stopping students from taking tests, and aggravating mental health issues. In reality,web security  is one of the most egregious instances of employing AI to automate ineffective pedagogical techniques rather than to build creative alternatives.
\section{Challenges}
Education 5.0 is a concept that refers to the next generation of education, which is characterized by a more personalized, learner-centric approach that utilizes technology to enhance the learning experience. It is based on the idea that education should be flexible, adaptable, and responsive to the needs of individual learners \cite{elayyan2021future}, \cite{huang2021emergence}. Some of the major challenges of Education 5.0 are explained below.

One of the major challenges for Education 5.0 is implementation cost. To implement a personalized, technology-enabled education system, schools and colleges must invest in new technologies such as learning management systems, digital content, and hardware \cite{al2020survey}, \cite{corbett2020connectivism}. This can be a huge financial burden for schools and colleges with limited budgets. The cost of maintaining and upgrading these technologies over time can also be high. This can make it difficult for schools and universities to implement Education 5.0 fully. This is especially true in developing countries with limited resources.

Another major challenge for Education 5.0 is the lack of teacher training \cite{ayyoubzadeh2023remote}. To effectively implement a personalized, technology-enabled education system, teachers must be trained to use new technologies and teaching methods. This can be a major challenge, especially in developing countries with limited resources. Additionally, many teachers are resistant to change and may not be comfortable using new technology in the classroom \cite{gupta2023teachers}. Without proper training, a teacher may not be able to effectively use the core technology and teaching methods of her Education 5.0.

The digital divide is also a big challenge for Education 5.0 \cite{gan2022iot}, \cite{kumar2022iot}. Not all students have access to the same technologies and resources. This can create a digital divide where some students have access to the latest technology and resources while others do not. This could further widen the gap between students from different socioeconomic backgrounds. This can make it difficult for educators to provide personalized, technology-enabled education for all students, regardless of background.
Limited resources for content creation are also a challenge in implementing Education 5.0 \cite{ejaz2022real}. Developing interactive and personalized learning materials can be a time-consuming and expensive process and can be a barrier for schools and colleges with limited resources. Additionally, the lack of standardization in the technologies and platforms used by educational institutions can make resource sharing and exchange difficult.

Privacy and security are also major challenges for Education 5.0 \cite{reidenberg2018achieving}. Privacy and security concerns are becoming more and more important as personal and sensitive information is shared and stored online. This can be a major challenge for schools and colleges that need to ensure the safety and security of student data. The threat of cyberattacks and data breaches is becoming more prevalent, and schools and colleges must take the necessary steps to protect student data \cite{torres2023pre}.

Another challenge is the limited research on the effectiveness and impact of Education 5.0 \cite{Shah2021}. As a relatively new concept, research on its efficacy and impact is still limited. This makes it difficult for educators and policymakers to make informed decisions about its implementation. Moreover, without proper research, it is difficult to determine the best way to implement Education 5.0 and measure its success.

One of the hot technology nowadays is Metaverse \cite{abraham2023study}. Implementing the Metaverse in education requires a significant investment in technical infrastructure, including high-speed internet access, powerful servers, and hardware devices such as VR/AR headsets. Developing educational content that is suitable for the Metaverse can be a complex and time-consuming task. Teachers and educators need to be trained on how to create and deliver content in a virtual environment. Not all students have access to the necessary technology and devices to participate in the Metaverse. This can create a digital divide, where some students have an advantage over others. The Metaverse raises concerns about security and privacy, as personal data and sensitive information may be at risk of being hacked or stolen. 

The Metaverse is a complex and rapidly evolving technology, and there is a lack of standardization and interoperability between different platforms and devices \cite{lin2022metaverse}, \cite{hyun2023study}. The Metaverse can be difficult to scale, as the number of users and devices increases \cite{lin2022metaverse}, \cite{Shah2021}. This can lead to technical challenges such as latency, bandwidth constraints, and server overload \cite{lin2022metaverse}, \cite{hyun2023study}. Providing an intuitive and seamless user experience is crucial to the success of the Metaverse. The Metaverse should be easily navigable and engaging, with minimal lag or glitches.

Finally, balancing online and offline learning is also a challenge for Education 5.0 \cite{lee2022offline}. It can be difficult to find the right balance between online and offline learning, as too much online learning can lead to isolation and lack of social interaction, while too much offline learning can limit the use of technology and personalization.

\begin{table}[h!]
    \centering
    \caption{Summary of challenges in the realization of Education 5.0 and open questions.}
    \begin{tabular}{p{1cm}|p{3.2cm}|p{4.5cm}|p{7.5cm}}
         \cline{1-4}
         \textbf{Ref}. & \textbf{Challenges} & \textbf{Possible Solutions} & \textbf{Open Questions} \\ \cline{1-4}
         \cite{kumar2022iot},\cite{gan2022iot} & Implementation cost & Investing in new technologies & How to implement Education 5.0 with limited budgets? How to maintain and upgrade the technologies over time? \\ \cline{1-4}
         \cite{hyun2023study} & Lack of teacher training & Providing teacher training & How to provide teacher training in developing countries with limited resources? How to overcome resistance to change among teachers? \\ \cline{1-4}

        \cite{lin2022metaverse} &  Digital divide & Providing equal access to technology and resources for all students & How to address the digital divide in education? How to ensure that all students have access to the necessary technology and resources? \\ \cline{1-4}

        \cite{Shah2021} & Limited resources for content creation & Developing interactive and personalized learning materials & How to make resource sharing and exchange easier? How to develop interactive and personalized learning materials with limited resources? \\ \cline{1-4}

        \cite{corbett2020connectivism} & High implementation cost & Investing in new technologies & How to implement Education 5.0 with limited budgets? How to maintain and upgrade the technologies over time? \\ \cline{1-4}

        \cite{ayyoubzadeh2023remote} & Lack of teacher training & Providing teacher training &  How to provide teacher training in developing countries? How to provide teacher training in developing countries with limited resources? How to overcome resistance to change among teachers? \\ \cline{1-4}

        \cite{gupta2023teachers} & Limited research on the effectiveness and impact of Education 5.0 & Conducting more research & How to determine the best way to implement Education 5.0? How to measure its success? \\ \cline{1-4}

        \cite{lin2022metaverse} & Privacy and security & Ensuring the safety and security of student data & How to protect student data from cyberattacks and data breaches? How to address privacy concerns in a technology-enabled education system? \\ \cline{1-4}

        \cite{lin2022metaverse} & Implementing the Metaverse in education & Investing in technical infrastructure and training teachers & How to provide an intuitive and seamless user experience? How to overcome the digital divide in the Metaverse? How to ensure the security and privacy of personal data in the Metaverse? \\ \cline{1-4}

        \cite{lee2022offline} & Balancing online and offline learning & Finding the right balance between online and offline learning & How to ensure social interaction while providing online learning? How to use technology and personalization in offline learning? \\ \cline{1-4}

        \cite{Shah2021} & Limited resources for content creation & Developing interactive and personalized learning materials & How to make resource sharing and exchange easier? How to develop interactive and personalized learning materials with limited resources? \\ \cline{1-4}

        \cite{hyun2023study} & Privacy and security & Ensuring the safety and security of student data & How to protect student data from cyberattacks and data breaches? How to address privacy concerns in a technology-enabled education system? \\ \cline{1-4}

        \cite{lin2022metaverse} & Implementing the Metaverse in education & Investing in technical infrastructure and training teachers & How to provide an intuitive and seamless user experience? How to overcome the digital divide in the Metaverse? How to ensure the security and privacy of personal data in the Metaverse? \\ \cline{1-4}

        \cite{lee2022offline} & Balancing online and offline learning & Finding the right balance between online and offline learning & How to ensure social interaction while providing online learning? How to use technology and personalization in offline learning? \\ \cline{1-4}

    \end{tabular}
    
    \label{tab:challenges}
\end{table}

\subsection{Summary and Insights}

In conclusion, Education 5.0 is a promising concept that can enhance the learning experience for students, but it also comes with a set of challenges. These challenges include implementation cost, lack of teacher training, digital divide, limited resources for content creation, privacy and security concerns and limited research on the effectiveness and impact of Education 5.0. To overcome these challenges, schools and colleges need to invest in new technologies, provide proper training for teachers, and ensure that all students have access to the necessary technology and resources. Additionally, proper research needs to be conducted to determine the best ways to implement Education 5.0 and measure its success. Furthermore, steps must be taken to ensure the privacy and security of student data, and to address any ethical concerns that may arise. Overall, Education 5.0 has the potential to revolutionize education, but it requires careful planning, implementation, and ongoing evaluation to ensure its success. Table \ref{tab:challenges} summarizes the different challenges that can potentially make barriers while implementing Education 5.0 systems. We identify the possible solution based on the current state-of-the-art and envision open questions and research direction to address that particular challenge.

\section{Conclusion}
In this survey, we have exhibited the concept of Education 5.0, a futuristic term for the next revolution in Education. The evolution of Education by the gradual integration of ICT and AI technologies which paved the way for Education 5.0 has been exhibited.  We proposed a conceptual 2D architecture to emphasize on the goal and requirements while the need for technology on the other end to complement the goal. The current state-of-the-art which lead to the idea of Education 5.0 is elucidated while some of the crucial challenges are identified to complete the study. We aim that this survey will cover a base for the realization of Education 5.0, as the next big wave in education.
\appendices


\normalem
\bibliographystyle{IEEEtran}
\bibliography{IEEEabrv,manuscript_r1}



\end{document}